\documentclass[aps,prd,amsmath,amsfonts,a4paper,reprint,preprintnumbers,twocolumn,nofootinbib,superscriptaddress,10pt]{revtex4-1}

\usepackage[utf8]{inputenc}

\usepackage{natbib}
\usepackage{amsmath}
\usepackage{amsfonts}
\usepackage{amssymb}
\usepackage{color}
\usepackage{bm}

\usepackage{extarrows}
\usepackage{accents}
\newcommand*{\bigdot}[1]{%
  \accentset{\mbox{\large\bfseries .}}{#1}}

\usepackage{graphicx}

\usepackage[labelfont={bf,scriptsize},textfont={scriptsize},justification=RaggedRight,format=hang]{caption,subfig}

\usepackage[T1]{fontenc}

\def\E{\mathcal{E}}

\newcommand{\indq}[1]{c_{#1}d_{#1},e_{#1}f_{#1}}

\newcommand{\lie}{\mathsterling\!}

\newcommand{\pd}{\partial}

\newcommand{\wt}[1]{\widetilde{#1}}

\newcommand{\red}{\textcolor{red}}

\def\({\left(}
\def\){\right)}
\def\<{\left<}
\def\>{\right>}
\newcommand{\be}{\begin{equation}}
\newcommand{\ee}{\end{equation}}
\newcommand{\bmat}{\left(\begin{matrix}}
\newcommand{\emat}{\end{matrix}\right)}
\newcommand{\nn}{\nonumber\\}

\newcommand{\lagr}{\mathcal{L}}

\usepackage{cancel}

\definecolor{dullpurple}{rgb}{0.431,0.188,0.534}
\definecolor{darkgreen}{rgb}{0.133,0.545,0.133}

\RequirePackage[colorlinks=true
,urlcolor=dullpurple
,anchorcolor=blue
,citecolor=dullpurple
,filecolor=blue
,linkcolor=darkgreen
,menucolor=blue
,pagecolor=blue
,linktocpage=true
,pdfproducer=medialab
]{hyperref}

\begin{document}

\title{On the uniqueness of ghost-free special gravity}

   \author{Dong Bai}
   \email{dbai@itp.ac.cn}
\affiliation{Key Laboratory of Theoretical Physics, Institute of Theoretical Physics,\\Chinese Academy of Sciences, Beijing 100190, China}\affiliation{School of Physical Sciences, University of Chinese Academy of Sciences,\\No.19A Yuquan Road, Beijing 100049, China}
	
    \author{Yu-Hang Xing}
    \email{xingyh@itp.ac.cn}
\affiliation{Key Laboratory of Theoretical Physics, Institute of Theoretical Physics,\\Chinese Academy of Sciences, Beijing 100190, China}\affiliation{School of Physical Sciences, University of Chinese Academy of Sciences,\\No.19A Yuquan Road, Beijing 100049, China}

\begin{abstract}

Special gravity refers to interacting theories of massless gravitons in Minkowski space-time which are invariant under the abelian gauge invariance $h_{ab}\rightarrow h_{ab}+\pd_{(a}\chi_{b)}$ only. In this article we determine the most general form of special gravity free of Ostrogradski ghosts, meaning its equation of motion is of at most second order. Together with the recent works, this result could be helpful in formulating proofs of General Relativity as the unique physical theory of self-interacting massless gravitons. We also study how to construct gauge invariant couplings to matter fields.

\end{abstract}
	
\maketitle

\section{Introduction}
	\label{sec:introduction}
	
\subsection{Background}
Possibilities of modifying General Relativity (GR) have been continuously explored ever since its appearance. Some attempts are of phenomenological interest, while others are of pure theoretical interest. This article belongs to the latter case at least.

There have been long-lasting attempts to show that GR is the unique physical theory of interacting massless gravitons from fundamental principles of special relativity and quantum mechanics \cite{Fierz:1939ix,Gupta:1954zz,Kraichnan:1955zz,Feynman:1996kb,Deser:1969wk,Wald:1986bj,Deser:1987uk,Padmanabhan:2004xk,Deser:2009fq,Butcher:2009ta,Barcelo:2014mua}. In many ``proofs'' of the GR uniqueness, typically one needs not only well-established principles such as Lorentz invariance and unitarity, but also some technical assumptions such as minimal couplings with only two derivatives, universal matter couplings inspired by the classical equivalence principle\footnote{Recently it is shown that the classical equivalence principle could be violated by quantum effects \cite{Bjerrum-Bohr:2014zsa,Bjerrum-Bohr:2016hpa,Bai:2016ivl}. As a result, it is questionable to have the classical equivalence principle as a physical input assumption.}, etc,  which are physically less robust than various fundamental principles. One of the most general results of this style with only few additional assumptions is given by Wald, stating that given the technical assumption that the equations of motion could be derived from action principles, the gauge transformation of free massless graviton has only one possible non-linear extension, that is, diffeomorphism transformation acting on a symmetric rank-2 covariant tensor \cite{Wald:1986bj}. In other words, if a weakly coupled\footnote{Here by ``weakly coupled'' we mean the non-linear theory has no more degrees of freedom than the free theory.} theory were to support a Minkowski vacuum with massless-graviton excitations around it, then either the theory could be rewritten in the geometric language using a metric as the dynamical field, or it remains to be perturbative field theory in Minkowski spacetime enjoying linear abelian gauge invariance only. Temporarily, we shall call the former and latter possibilities as theories of the first and second type respectively.

Theories of the first type have been studied extensively a long time ago. One of the main achievements along this direction is given by Ref.~\cite{Lovelock:1971yv}, which shows that the only theories enjoying second-order equations of motion are GR (in 4 dimensions) and Lovelock gravity (in $D>4$ dimensions). This result is known as the Lovelock theorem in the literature. 

On the other hand, theories of the second type are relatively less examined. Compared to theories of the first type, these theories are defined rigorously in the Minkowski spacetime and look more like the traditional quantum field theories. Wald himself did write down some examples of such theories for illustrative purposes, which obviously involve higher derivatives in equations of motion \cite{Wald:1986bj}. The first appearance of ghost-free interactions is done by Ref.~\cite{Hinterbichler:2013eza}, where the authors search for ghost-free kinetic modifications to Lorentz invariant massive gravity \cite{deRham:2010ik,deRham:2010kj,Hinterbichler:2011tt,deRham:2014zqa,deRham:2016nuf}. It turns out some of the terms they obtained are valid for massless gravitons also. In a previous work \cite{Bai:2016hui} we named these theories enjoying only the abelian gauge invariance as {\it special gravity}, to emphasize special covariance. {\it Abelian gravity} could also be a good name but it has been adopted by another theory \cite{Broda:2007qw}. In the same work, we studied the three-point vertices of special gravity using various modern techniques developed by the particle physics community, such as the spinor-helicity formalism, asymptotic causality \cite{Gao:2000ga,Camanho:2014apa}, etc. It has been shown already in Ref.~\cite{Camanho:2014apa} that asymptotic causality could be helpful in picking out GR from modified gravity theories of the Lovelock type. In Ref.~\cite{Bai:2016hui}, we show further that the same principle could also be helpful in eliminating special gravity as physical theories at the fundamental level, as the three-point vertices violate explicitly asymptotic causality by themselves. At present, the asymptotic-causality arguments are limited to the three-point vertices, and new insights are needed to extend the analysis to higher-point vertices. Anyway, together with Ref.~\cite{Wald:1986bj,Camanho:2014apa} we figure out that the causality principle could play an important role in formulating the uniqueness of GR, helping to eliminate non-GR theories of both the first type and second type and allowing one to replace various ad hoc technical assumptions with more physical ones (see also Ref.~\cite{Hertzberg:2016djj,Hertzberg:2017abn} for relevant discussions). 

It is worthwhile to notice that although special gravity may not be a theory of fundamental interactions by themselves due to potential violations of asymptotic causality, there could be some applications in condensed matter physics. It is figured out by Ref.~\cite{Zhang:2001xs,Gu:2009jh,Zaanen:2011hm} theoretically that it is possible to construct a condensed matter system which contains emergent relativistic massless graviton excitations at the long distance. The effective descriptions of these emergent gravitons are Lorentz invariant (with an effective speed of light). The authors of Ref.~\cite{Zhang:2001xs,Gu:2009jh,Zaanen:2011hm} attempt to claim that the interaction of these massless graviton excitations would serve as a low-speed-of-light version of GR. However, it is known by the particle physics community that this is not the case, as condensed matter systems typically contain local degrees of freedom, while GR does not. As a result, we propose special gravity to be a better candidate, which is actually nothing but field theories defined in the flat spacetime and contains local observables just like the traditional quantum field theories. The aforementioned violations of asymptotic causality are less relevant in these cases, as the short-distance descriptions of the condensed matter systems are typically non-relativistic, and thus there is no relativistic notion of causality at the fundamental level. If these condensed matter systems were realized in the lab \cite{Price:2015qca,Ozawa:2016htj}, it would be possible to see special gravity in the real nature. Also, there could be other applications in the studies of massive gravity (see Ref.~\cite{Bai:2016hui} for details).

With all these in mind, it is meaningful to continue theoretical studies of special gravity. One of the questions unanswered by Ref.~\cite{Bai:2016hui} is what is the most general ghost-free special gravity. This question is answered by the present article. We have tried to be mathematically rigorous instead of making ambiguous statements.

\subsection{Main result}
Under the following conditions
\begin{itemize}
	\item The equations of motion are at most second-order,
	\item The theory is {\it Lagrangian} in nature, meaning that the equations of motion are derivable from an action principle,
\end{itemize}
the only terms of self-interacting massless gravitons in Minkowski spacetime that enjoy abelian gauge invariance are given by:
\begin{equation}
\lagr^{(n)}=\, h_{[a}^{a} \partial_{a_1}\partial^{a_1}h_{b_1}^{b_1}\cdots  \partial_{a_n}\partial^{a_n}h_{b_n]}^{b_n},
\end{equation}
where $h_{ab}$ denotes the graviton field. The $n$-th term is nontrivial only in dimensions $D>=2n+1$.

\subsection{Notations and terminology}

Throughout this paper we make use of the following definitions of notations: comma (,) means space-time partial differentiation, while semicolon (;) is dedicated to local functions of tensor fields and means partial differentiation with respect to relevant tensor fields or tensor fields with the space-time derivatives. For example
\begin{subequations}
\begin{align}
h_{ab,cd}\ &\bigdot{=}\ \pd_d\pd_c h_{ab},	\\
E^{ab;cd,ef}\ &\bigdot{=}\ \frac{\pd E^{ab}}{\pd h_{cd,ef}},	
\end{align}
\end{subequations}
which agrees with that of Ref.~\cite{Deffayet:2013tca}. 

Also, the parenthesis $T_{(abcd)efg}$ means symmetrization while the bracket $T_{[abcd]efg}$ means anti-symmetrization. Indices are freely raised and lowered by the flat metric $\eta_{ab}$

We also have to clarify that by {\it Lorentz invariant} we really mean {\it Poincare invariant}. This applies almost everywhere in this article. Lastly, keep in mind that {\it gauge invariance}, {\it linear gauge invariance}, {\it abelian gauge invariance} all mean the same thing in this article. We sometimes use the ancient word {\it concomitant} to mean a tensorial expression which is constructed locally from several tensors in accordance with D. Lovelock and G. Horndeski.

\section{The proof}

\subsection{Gauge invariance}
\label{sec:gaugeinv}

In this section we'll derive the constraints that gauge invariance puts on the equations of motion  $E^{ab}$ which is required to be a local function of $h_{ab}$, $h_{ab,c}$ and $h_{ab,cd}$, and {\it manifestly} gauge invariant.

Let $\wt{E}^{ab}$ denote $E^{ab}$ with $h_{ab}$ replaced by $h_{ab}+\xi_{(a,b)}$, where the gauge transformation parameter is denoted by $\xi$. Note that in general $\wt{E}^{ab}$ is a function of both $h,\pd h, \pd\pd h$ and $\pd\xi,\pd\pd\xi,\pd\pd\pd\xi$. But because of gauge invariance, $\wt{E}^{ab}$ has to be independent of all the latter arguments, which is possible if and only if
\begin{subequations}\label{eq:cc}
\begin{align}
0=\frac{\pd\wt{E}^{ab}}{\pd \xi_{c,d}} &= E^{ab;(cd)}=E^{ab;cd}	\label{eq:cc1},\\
0=\frac{\pd\wt{E}^{ab}}{\pd \xi_{c,de}} &= E^{ab;c(d,e)}	\label{eq:cc2},\\
0=\frac{\pd\wt{E}^{ab}}{\pd \xi_{c,def}} &= E^{ab;c(d,ef)}	\label{eq:cc3}.
\end{align}
\end{subequations}

Note that the present case is much simpler than the generally covariant (Lovelock) case \cite{Lovelock:1971yv} in that different orders of derivative of the gauge parameter field don't compensate each other. Here we have decoupled constraints.

Eq.~\eqref{eq:cc1} says that $E^{ab}$ simply can't depend on $h$; Eq.~\eqref{eq:cc2} says that $E^{ab;cd,e}$ is antisymmetric in $d,e$, but it is by definition symmetric in $c,d$. The incompatibility of those two symmetries are well known thus $E^{ab;cd,e}$ vanishes, so $E^{ab}$ can't depend on $\pd h$ either. We henceforth drop any dependences on the zeroth and first spacetime derivatives of $h_{ab}$, which greatly simplifies the analysis.

\subsection{Integrability (Lagrangianity)}
\label{sec:integrability}

In this section we'll derive the condition under which the equations of motion are derivable from an action principle. Suppose the equations of motion are derivable from an action functional $S$, then for commutativity of functional derivatives
\begin{align}
0 & \equiv \left[\frac{\delta}{\delta h_{ab}(x)},\frac{\delta}{\delta h_{cd}(y)}\right]S	\nn
& = \frac{\delta}{\delta h_{ab}(x)} E^{cd}(y) - \frac{\delta}{\delta h_{cd}(y)} E^{ab}(x).
\end{align}
This is what we call {\it integrability} (or more fancifully {\it Lagrangianity}) condition. It's a necessary and sufficient condition. Taking further Eq.~\eqref{eq:cc1} and \eqref{eq:cc2} into account, we get
\begin{align}
0\equiv & E^{cd;ab,ef}(y)\pd^y_e\pd^y_f\delta^D(x-y) \nn 
& - E^{ab;cd,ef}(x)\pd^x_e\pd^x_f\delta^D(x-y).
\end{align}
This expression should be understood \textit{distributionally}. To extract information we multiply it with two test functions $f(x)$ and $g(y)$ and do the integration. We have
\begin{align}
& \int d^Dxd^Dy\ f(x)g(y)E^{cd;ab,ef}(y)\pd^y_e\pd^y_f\delta^D(x-y) 	\nn
= & \int d^Dxd^Dy\ \pd^y_e\pd^y_f\left[ g(y)E^{cd;ab,ef}(y) \right] f(x)\delta^D(x-y)	\nn
= & \int d^Dy\ \pd^y_e\pd^y_f\left[ g(y)E^{cd;ab,ef}(y) \right] f(y),
\end{align}
\begin{align}
& \int d^Dxd^Dy\ f(x)g(y)E^{ab;cd,ef}(x)\pd^x_e\pd^x_f\delta^D(x-y)	\nn
= & \int d^Dx\ \pd^x_e\pd^x_f\left[ f(x)E^{ab;cd,ef}(x) \right] g(x)	\nn
= & \int d^Dx\ f(x)E^{ab;cd,ef}(x) \pd^x_e\pd^x_f g(x).
\end{align}
After some simplifications and taking into account that the identity holds for any test function $f(x)$, we get
\begin{align}
0\equiv & \pd_e\pd_f\left[ g(x)E^{cd;ab,ef}(x) \right]  - E^{ab;cd,ef}(x) \pd_e\pd_f g(x)		\nn
=& \pd_e\pd_f g(x) \left(E^{cd;ab,ef} - E^{ab;cd,ef}\right) 	\nn
& + 2 \pd_e g(x)\pd_f E^{cd;ab,ef} + g(x) \pd_e\pd_f E^{cd;ab,ef}.
\end{align}
This in turn holds for arbitrary test function $g(x)$, thus
\begin{subequations}\label{eq:integrability}
\begin{align}
E^{cd;ab,ef} - E^{ab;cd,ef}=0,	\label{eq:integrability1}\\
\pd_f E^{cd;ab,ef}=0,	\\
\pd_e\pd_f E^{cd;ab,ef}=0.	
\end{align}
\end{subequations}
The first one says $E^{ab;cd,ef}$ is symmetric under the exchange of $ab \leftrightarrow cd$; the third one is weaker than the second, which says
\begin{align}
0&=\pd_f E^{cd;ab,ef}	\nn
&=E^{cd;ab,ef;gh,pq}h_{gh,pqf}.
\end{align}
Therefore
\begin{align}
E^{cd;ab,e(f|;gh,|pq)}=0	\label{eq:integrability2}.
\end{align}

\subsection{General form of $E^{ab}$}

Here we summarize the properties of $E^{ab;cd,ef}$ :
\begin{subequations}
\begin{align}
& E^{ab;c(d,ef)} = 0,	\label{eq:summary1}\\
& E^{cd;ab,ef} =  E^{ab;cd,ef}, \\
& E^{ab;cd,ef} = E^{ab;ef,cd},		\\
& E^{ab;\cdots;cd,ef;\cdots;gh,pq;\cdots} = E^{ab;\cdots;gh,pq;\cdots;cd,ef;\cdots}.
\end{align}
\end{subequations}

The first two are just \eqref{eq:cc3} and \eqref{eq:integrability1}; the third one is derivable from the first one; the last one is due to commutativity of partial derivatives ``;''. We've discarded \eqref{eq:integrability2} because it could be derived from the above four properties.

Put in words, the index pairs in $E^{ab;\cdots}$ satisfy {\it Property S} defined in \cite{Lovelock:1969uni}. For such a set of index pairs, whenever three of the indices coincide, the expression vanishes. This is because one can always bring any three identical indices into a cyclic group by repetitive use of the cyclic identity \eqref{eq:summary1} which holds for any two pairs of indices thanks to the symmetry properties. So there's an upper bound on number $k$ of partial derivatives with respect to $h_{ab,cd}$ in a given dimension $D$, namely
\begin{equation}\label{eq:dimbound}
4k+2\le 2D.
\end{equation}
Otherwise, there would always be three identical indices. 

Thus $E^{ab}$ has the following general form
\begin{align}
& \E^{ab}+\E^{ab;\indq{1}}h_{\indq{1}}+		\nn
& \E^{ab;\indq{1};\indq{2}}h_{\indq{1}}h_{\indq{2}}+\cdots+	\nn
& \E^{ab;\indq{1};\cdots;\indq{K}}h_{\indq{1}}\cdots h_{\indq{K}},
\end{align}
where $K=[\frac{D-1}{2}]$. The $\E$'s obviously enjoy Property S and are Lorentz invariant tensors. Appendix~\ref{sec:lovelocktheorem3} shows that the $\E$'s are determined to the unique form (up to a constant factor)
\begin{equation}
\eta^{[a}_b\eta^{c_1}_{d_1}\eta^{e_1}_{f_1}\cdots \eta^{c_K}_{d_K}\eta^{e_K]}_{f_K},
\end{equation}
where we temporarily lowered half of the indices for brevity of illustration.

\subsection{The Lagrangian}

We only have to find one Lagrangian that correctly gives rise to the equations of motion, because all Lagrangians giving rise to the same equations of motion differ only by a boundary term. The Lagrangian we choose is
\begin{equation}
\lagr^{(n)}=\, h_{[a}^{a} \partial_{a_1}\partial^{a_1}h_{b_1}^{b_1}\cdots  \partial_{a_n}\partial^{a_n}h_{b_n]}^{b_n}.
\end{equation}
It's easy to see this correctly reproduces the desired equations of motion upon variation.

This family of Lagrangians happen to be those ``pseudo-linear'' Lagrangians corresponding to the Lovelock terms, which was already studied by \cite{Hinterbichler:2013eza,Gao:2014jja}. The invariance of such terms under linear gauge transformation was already pointed out by those authors. There's also a general proof provided in the appendix of \cite{Bai:2016hui}. Now we see that these happen to be the {\it only} ghost-free gauge invariant Lagrangians.

\section{Discussions}

\subsection{The field strength tensor}

Making use of Property S of the $\E$'s, we can cast the equations of motion into a form which depends on $h_{cd,ef}$ only through the combination $$h_{d[c,e]f}-h_{f[c,e]d},$$ which is just $R^{(1)}_{cedf}[\eta+h]$, the first order expansion of the Riemann curvature tensor, and is manifestly gauge invariant.

Interestingly enough there are no way to do the same to the corresponding Lagrangian, which is at best gauge invariant up to boundary terms.

\subsection{Relation with Deser's iterative procedure}
There's a textbook procedure developed long ago by Ref.~\cite{Deser:1969wk}, where GR could be brought up iteratively out of a free massless graviton Lagrangian with additional couplings to the matter energy-momentum tensor. The general procedure is
\begin{itemize}
	\item Start with the free Lagrangian $h_{[a}^b\pd_c\pd^d h_{e]}^f$.
	\item Couple it to the energy-momentum tensor of some previously isolated matter sector through $h_{ab}T^{ab}$. Since a conserved $T^{ab}$ couples only to the transversal part of $h_{ab}$, this was expected to preserve gauge invariance.
	\item But once coupled, $T^{ab}$ is no longer conserved by itself, which in turn excites the longitudinal component of $h_{ab}$, destroying gauge invariance. To compensate for the non-conservation we try to add the energy-momentum tensor of the graviton itself to $T^{ab}$, which gives rise to a self-coupling term of $h_{ab}$.
	\item But this self-coupling also contributes a higher order term to $T^{ab}$, and eventually we find ourselves doing this for an infinite number of times, and find out the terms that we add agree with the flat-space expansion of GR order by order.
\end{itemize}

In the second step there's the assumption that the graviton has to couple to energy-momentum tensor. This is a natural assumption which seems too reasonable to drop. But what we have to say is that it is the removal of this very assumption that gives rise to many interesting possibilities, like the $\Lambda_3$ decoupling limit of massive gravity \cite{deRham:2010ik,deRham:2010kj}, where $h_{ab}$ couples to a symmetric tensor $\chi^{ab}$ which is not the energy-momentum tensor but is {\it identically} conserved, meaning $\pd_a\chi^{ab}=0$ holds without any external help. Below is an example
\begin{equation}\label{eq:decouplim}
h^{ab}(\pd_a\pd_b\phi-\eta_{ab}\pd^2\phi).
\end{equation}
 If in the second step we were to add this term instead of $h_{ab}T^{ab}$, there would be nothing to do further, the theory is already complete.

In special gravity the situation is similar in the sense that the equations of motion satisfies $\pd_aE^{ab}=0$ identically (off-shell), which is a necessary condition for gauge invariance.

\subsection{Coupling to matter fields}
Eq~\eqref{eq:decouplim} is a working example of healthy coupling of the special graviton with a scalar field. We now describe an algorithm to construct more of such couplings.

For this purpose, note that the $\pd_a\pd_b\phi-\eta_{ab}\pd^2\phi$ can be derived by functionally differentiating the following action $\int \sqrt{-g}R[g]\phi$ with respect to $g_{ab}$, and then in the obtained expression setting $g_{ab}=\eta_{ab}$. This gives us a hint of how to generalize.

In fact, any action of the form $S[g_{ab},\Phi]$ (where $\Phi$ denotes a collection of tensor fields) which vanishes when $g_{ab}=\eta_{ab}$, could give rise to an (identically conserved) symmetric tensor upon variating with respect to $g_{ab}$ and then setting $g_{ab}=\eta_{ab}$, since
\begin{align}
0 &= \frac{d}{dt}S[\phi^*_{Xt}g,\phi^*_{Xt}\Phi]	\nn
&=\int 2\frac{\delta S}{\delta g_{ab}}\nabla_{(a}X_{b)}+E_{\Phi}\cdot\lie_X\Phi,
\end{align}
for any test vector field $X$ vanishing on the space-time boundary. Now set $g_{ab}=\eta_{ab}$, since the action $S$ vanishes, $E_{\Phi}=\frac{\delta S}{\delta\Phi}$ is zero by definition. We get that $\pd_a\chi^{ab}=0$ holds identically, where we have defined
$$\chi^{ab}=\left.\frac{\delta S}{\delta g_{ab}}\right|_{g_{ab}=\eta_{ab}}.$$

Below is an example. Let's borrow $\int \sqrt{-g}G_{ab}\nabla^a\phi\nabla^b\phi$ from the Horndeski family, where $G_{ab}$ is the Einstein tensor. It's a good choice since the resulting $\chi^{ab}$ would contain no higher order derivatives. Then we could obtain the gauge invariant coupling
\begin{align}
&-h_{ab}\pd_c\pd^a\phi\pd^c\pd^b\phi -\frac{1}{2}h_a^a(\pd^2\phi)^2 	\nn
&+\frac{1}{2}h_a^a\pd_c\pd_d\phi\pd^c\pd^d\phi + h_{ab}\pd^a\pd^b\phi\pd^2\phi.	\nn
\end{align}

The procedure described above is a special case of the ``pseudo-linear'' construction. This could be seen by working out another example: $\int \sqrt{-g}\mathcal{G}\phi$, where $\mathcal{G}$ is the 4-dimensional Euler Density. This is a Horndeski term which vanishes for both zeroth and first order in $h_{ab}=g_{ab}-\eta_{ab}$, thus the second order expansion in $h_{ab}$
$$\phi\partial_{[a}\partial^{a}h_{b}^{b}\partial_{c}\partial^{c}h_{d]}^{d}.
$$
would be a gauge invariant $h$-$h$-$\phi$ vertex. Gauge invariant vertices involving more graviton legs could be obtained in this way.

\subsection{Interaction between multiple special gravitons}
\newcommand{\ha}[3]{h_{\ \ \ #2}^{(#1)#3}}
In the main part we only dealt with self-interaction of a single massless graviton, but the Lagrangian is readily generalizable to multiple fields:
\begin{equation}
\lagr^{(n)}=\, C_{\alpha_1\alpha_2\cdots \alpha_{n+1}}\ha{\alpha_1}{[a}{a} \partial_{a_1}\partial^{a_1}\ha{\alpha_2}{b_1}{b_1}\cdots  \partial_{a_n}\partial^{a_n}\ha{\alpha_{n+1}}{b_n]}{b_n}.
\end{equation}
where $\alpha$'s are internal indices and $C_{\alpha_1\alpha_2\cdots \alpha_{n+1}}$ is some arbitrary coefficient with restrictions coming only from the internal symmetries. One verifies with ease that the equations of motion are no more than second order. This is in sharp contrast with GR, where two gravitons won't interact with each other easily.

\section{summary}
In this paper, we construct the most general form of ghost-free special gravity, and discuss its relation to the iterative construction procedure of GR. We also develop a routine to seek for gauge invariant couplings between special gravitons and matter fields.

\acknowledgements{We would like to thank Prof. Qing-Guo Huang for the many useful discussions. We would also like to thank the authors of Ref.~\cite{Nutma:2013zea} for developing the excellent Mathematica packages xTras which we have used extensively for symbolic calculations.
}

\appendix
\section{The coefficients $\E$'s}
\label{sec:lovelocktheorem3}
\newtheorem{Claim}{Claim}
We are to determine the most general rank-$(2L)$ contravariant tensor which is Poincare invariant and enjoys Property S. Note that we are just one claim away from Theorem 3 of Ref.~\cite{Lovelock:1971yv}:

\begin{Claim}
Any Poincare invariant tensor can be expressed by a local tensorial expression of the flat metric $\eta$, that is 
$$T^{a_1a_2\cdots} = T^{a_1a_2\cdots}(\eta).$$
\end{Claim}
{\bf Proof} \ Firstly, since it's Poincare invariant, it must be a concomitant of $\eta_{ab}$, $\pd\eta$, $\pd\pd\eta$, etc. Secondly, the partial derivatives are simple to rule out since we are free to choose a Minkowski frame in which all partial derivatives of $\eta$ vanish, then the expression reduces to an expression independent of the derivatives but is {\it by itself} tensorial, meaning it could be used in any frame.

The rest of this appendix is just the Lovelock's theorem put in modern notations. The original proofs \cite{Lovelock:1971yv,Lovelock:1969uni} involve the contents of a series of papers and are too ancient to read. For this reason we recommend continuing with this appendix instead of searching through those papers.

\begin{Claim}
Any local tensorial expression $T^{a_1a_2\cdots}(g_{ab})$ where $g$ is a metric field satisfies the following identity 
\begin{equation}\label{eq:Tsymab}
\sum_{k=1}^{\cdots}T^{\cdots a_{k-1}aa_{k+1}\cdots}g^{a_kb}=\sum_{k=1}^{\cdots}T^{\cdots a_{k-1}ba_{k+1}\cdots}g^{a_ka}.
\end{equation}
\end{Claim}
{\bf Proof} \ The mathematical form of tensoriality says
$$\phi^*(T(g))=T(\phi^*g),$$
where $\phi$ is an arbitrary diffeomorphism which we now choose to be generated from a vector field $X$ and parametrized by $t$,
$$\phi^*_{Xt}(T(g))=T(\phi^*_{Xt}g).$$
Differentiating both sides with respect to $t$ and then setting $t=0$, we get
$$(\lie_XT)^{a_1a_2\cdots}=\frac{\pd T^{a_1a_2\cdots}}{\pd g_{ab}}(\lie_Xg)_{ab}.$$
By direct calculation this becomes
\begin{equation*}
-\sum_{k=1}^{\cdots}T^{\cdots a_{k-1}aa_{k+1}\cdots}\nabla_a X^{a_k}=2\frac{\pd T^{a_1a_2\cdots}}{\pd g_{ab}}\nabla_{(a}X_{b)},
\end{equation*}
where $\nabla$ is the metric connection. Since this holds for arbitrary vector field $X$ we get
\begin{equation}\label{eq:tensoriality}
-\sum_{k=1}^{\cdots}T^{\cdots a_{k-1}aa_{k+1}\cdots}g^{a_kb}=\frac{\pd T^{a_1a_2\cdots}}{\pd g_{ab}}+\frac{\pd T^{a_1a_2\cdots}}{\pd g_{ba}}.
\end{equation}
Now the right hand side is manifestly symmetric in $a$ and $b$. The left hand side must also be so, which gives the desired result.
\begin{Claim}
If in addition $T^{a_1a_2\cdots}(g)$ has even number of indices grouped in pairs and they enjoy Property S, then the expression is determined up to a constant factor.
\end{Claim}
{\bf Proof} \ Say the number of indices is $2L$ and the spacetime dimension is $D$. Call it an {\it S-tensor} of rank-$L$ for short. Contracting Eq.~\eqref{eq:Tsymab} with $g_{a_1b}$ and using Property S, we get
\begin{align}\label{eq:dp1l}
&(D+1-L)\ T^{aa_2a_3\cdots}= 	\nn
&g_{a_1b}T^{a_1ba_3a_4\cdots}g^{aa_2}- \frac{1}{2}\sum_{p=3}^{2L}g_{a_1b}T^{a_1ba_3\cdots a_{p-1}a_2a_{p+1}\cdots}g^{aa_p}.
\end{align}
When $L>=D+1$, there are too many indices and $T$ vanishes by Property S (see Eq.~\eqref{eq:dimbound}). Thus we can safely put $L\le D$. Note the right hand side of Eq.~\eqref{eq:dp1l} is a combination of $g_{ab}$ and $g_{a_1b}T^{a_1ba_3a_4\cdots}$, with the latter to be an S-tensor of rank-(L-1). By recursive use of this equation we could eventually express the original S-tensor in terms of $g_{ab}$ and the scalar $g_{a_1a_2}g_{a_3a_4}\cdots T^{a_1a_2a_3a_4\cdots}$, with no undetermined coefficients.

It thus remains to prove that a scalar quantity constructed only from $g_{ab}$ must be a constant. But Eq.~\eqref{eq:tensoriality} with $T$ a scalar readily states the fact we want.

\bibliographystyle{JHEPmodplain}
\bibliography{references}

\providecommand{\href}[2]{#2}\begingroup\raggedright\begin{thebibliography}{10}

\bibitem{Fierz:1939ix}
M.~Fierz and W.~Pauli, {\it {On relativistic wave equations for particles of
  arbitrary spin in an electromagnetic field}},  {\sl Proc. Roy. Soc. Lond.}
  {\bf A173} (1939) 211--232,
  [\href{http://dx.doi.org/10.1098/rspa.1939.0140}{{\sf
  doi:10.1098/rspa.1939.0140}}].

\bibitem{Gupta:1954zz}
S.~N. Gupta, {\it {Gravitation and Electromagnetism}},  {\sl Phys. Rev.} {\bf
  96} (1954) 1683--1685, [\href{http://dx.doi.org/10.1103/PhysRev.96.1683}{{\sf
  doi:10.1103/PhysRev.96.1683}}].

\bibitem{Kraichnan:1955zz}
R.~H. Kraichnan, {\it {Special-Relativistic Derivation of Generally Covariant
  Gravitation Theory}},  {\sl Phys. Rev.} {\bf 98} (1955) 1118--1122,
  [\href{http://dx.doi.org/10.1103/PhysRev.98.1118}{{\sf
  doi:10.1103/PhysRev.98.1118}}].

\bibitem{Feynman:1996kb}
R.~P. Feynman, {\em {Feynman lectures on gravitation}}.
\newblock Reading, USA: Addison-Wesley (1995) 232 p. (The advanced book
  program), 1996.

\bibitem{Deser:1969wk}
S.~Deser, {\it {Selfinteraction and gauge invariance}},  {\sl Gen. Rel. Grav.}
  {\bf 1} (1970) 9--18, [\href{http://arxiv.org/abs/gr-qc/0411023}{{\sf
  arXiv:gr-qc/0411023}}], [\href{http://dx.doi.org/10.1007/BF00759198}{{\sf
  doi:10.1007/BF00759198}}].

\bibitem{Wald:1986bj}
R.~M. Wald, {\it {Spin-2 Fields and General Covariance}},  {\sl Phys. Rev.}
  {\bf D33} (1986) 3613,
  [\href{http://dx.doi.org/10.1103/PhysRevD.33.3613}{{\sf
  doi:10.1103/PhysRevD.33.3613}}].

\bibitem{Deser:1987uk}
S.~Deser, {\it {Gravity From Selfinteraction in a Curved Background}},  {\sl
  Class. Quant. Grav.} {\bf 4} (1987) L99,
  [\href{http://dx.doi.org/10.1088/0264-9381/4/4/006}{{\sf
  doi:10.1088/0264-9381/4/4/006}}].

\bibitem{Padmanabhan:2004xk}
T.~Padmanabhan, {\it {From gravitons to gravity: Myths and reality}},  {\sl
  Int. J. Mod. Phys.} {\bf D17} (2008) 367--398,
  [\href{http://arxiv.org/abs/gr-qc/0409089}{{\sf arXiv:gr-qc/0409089}}],
  [\href{http://dx.doi.org/10.1142/S0218271808012085}{{\sf
  doi:10.1142/S0218271808012085}}].

\bibitem{Deser:2009fq}
S.~Deser, {\it {Gravity from self-interaction redux}},  {\sl Gen. Rel. Grav.}
  {\bf 42} (2010) 641--646, [\href{http://arxiv.org/abs/0910.2975}{{\sf
  arXiv:0910.2975}}], [\href{http://dx.doi.org/10.1007/s10714-009-0912-9}{{\sf
  doi:10.1007/s10714-009-0912-9}}].

\bibitem{Butcher:2009ta}
L.~M. Butcher, M.~Hobson, and A.~Lasenby, {\it {Bootstrapping gravity: A
  Consistent approach to energy-momentum self-coupling}},  {\sl Phys. Rev.}
  {\bf D80} (2009) 084014, [\href{http://arxiv.org/abs/0906.0926}{{\sf
  arXiv:0906.0926}}], [\href{http://dx.doi.org/10.1103/PhysRevD.80.084014}{{\sf
  doi:10.1103/PhysRevD.80.084014}}].

\bibitem{Barcelo:2014mua}
C.~Barceló, R.~Carballo-Rubio, and L.~J. Garay, {\it {Unimodular gravity and
  general relativity from graviton self-interactions}},  {\sl Phys. Rev.} {\bf
  D89} (2014), no.~12 124019, [\href{http://arxiv.org/abs/1401.2941}{{\sf
  arXiv:1401.2941}}], [\href{http://dx.doi.org/10.1103/PhysRevD.89.124019}{{\sf
  doi:10.1103/PhysRevD.89.124019}}].

\bibitem{Bjerrum-Bohr:2014zsa}
N.~E.~J. Bjerrum-Bohr, J.~F. Donoghue, B.~R. Holstein, L.~Plante, and
  P.~Vanhove, {\it {Bending of Light in Quantum Gravity}},  {\sl Phys. Rev.
  Lett.} {\bf 114} (2015), no.~6 061301,
  [\href{http://arxiv.org/abs/1410.7590}{{\sf arXiv:1410.7590}}],
  [\href{http://dx.doi.org/10.1103/PhysRevLett.114.061301}{{\sf
  doi:10.1103/PhysRevLett.114.061301}}].

\bibitem{Bjerrum-Bohr:2016hpa}
N.~E.~J. Bjerrum-Bohr, J.~F. Donoghue, B.~R. Holstein, L.~Plante, and
  P.~Vanhove, {\it {Light-like Scattering in Quantum Gravity}},  {\sl JHEP}
  {\bf 11} (2016) 117, [\href{http://arxiv.org/abs/1609.07477}{{\sf
  arXiv:1609.07477}}], [\href{http://dx.doi.org/10.1007/JHEP11(2016)117}{{\sf
  doi:10.1007/JHEP11(2016)117}}].

\bibitem{Bai:2016ivl}
D.~Bai and Y.~Huang, {\it {More on the Bending of Light in Quantum Gravity}},
  \href{http://arxiv.org/abs/1612.07629}{{\sf arXiv:1612.07629}}.

\bibitem{Lovelock:1971yv}
D.~Lovelock, {\it {The Einstein tensor and its generalizations}},  {\sl J.
  Math. Phys.} {\bf 12} (1971) 498--501,
  [\href{http://dx.doi.org/10.1063/1.1665613}{{\sf doi:10.1063/1.1665613}}].

\bibitem{Hinterbichler:2013eza}
K.~Hinterbichler, {\it {Ghost-Free Derivative Interactions for a Massive
  Graviton}},  {\sl JHEP} {\bf 10} (2013) 102,
  [\href{http://arxiv.org/abs/1305.7227}{{\sf arXiv:1305.7227}}],
  [\href{http://dx.doi.org/10.1007/JHEP10(2013)102}{{\sf
  doi:10.1007/JHEP10(2013)102}}].

\bibitem{deRham:2010ik}
C.~de~Rham and G.~Gabadadze, {\it {Generalization of the Fierz-Pauli Action}},
  {\sl Phys. Rev.} {\bf D82} (2010) 044020,
  [\href{http://arxiv.org/abs/1007.0443}{{\sf arXiv:1007.0443}}],
  [\href{http://dx.doi.org/10.1103/PhysRevD.82.044020}{{\sf
  doi:10.1103/PhysRevD.82.044020}}].

\bibitem{deRham:2010kj}
C.~de~Rham, G.~Gabadadze, and A.~J. Tolley, {\it {Resummation of Massive
  Gravity}},  {\sl Phys. Rev. Lett.} {\bf 106} (2011) 231101,
  [\href{http://arxiv.org/abs/1011.1232}{{\sf arXiv:1011.1232}}],
  [\href{http://dx.doi.org/10.1103/PhysRevLett.106.231101}{{\sf
  doi:10.1103/PhysRevLett.106.231101}}].

\bibitem{Hinterbichler:2011tt}
K.~Hinterbichler, {\it {Theoretical Aspects of Massive Gravity}},  {\sl Rev.
  Mod. Phys.} {\bf 84} (2012) 671--710,
  [\href{http://arxiv.org/abs/1105.3735}{{\sf arXiv:1105.3735}}],
  [\href{http://dx.doi.org/10.1103/RevModPhys.84.671}{{\sf
  doi:10.1103/RevModPhys.84.671}}].

\bibitem{deRham:2014zqa}
C.~de~Rham, {\it {Massive Gravity}},  {\sl Living Rev. Rel.} {\bf 17} (2014) 7,
  [\href{http://arxiv.org/abs/1401.4173}{{\sf arXiv:1401.4173}}],
  [\href{http://dx.doi.org/10.12942/lrr-2014-7}{{\sf
  doi:10.12942/lrr-2014-7}}].

\bibitem{deRham:2016nuf}
C.~de~Rham, J.~T. Deskins, A.~J. Tolley, and S.-Y. Zhou, {\it {Graviton Mass
  Bounds}},  \href{http://arxiv.org/abs/1606.08462}{{\sf arXiv:1606.08462}}.

\bibitem{Bai:2016hui}
D.~Bai and Y.-H. Xing, {\it {Special Gravity as Alternatives for Interacting
  Massless Gravitons}},  \href{http://arxiv.org/abs/1610.00241}{{\sf
  arXiv:1610.00241}}.

\bibitem{Broda:2007qw}
B.~Broda, P.~Bronowski, M.~Ostrowski, and M.~Szanecki, {\it {Quantization of
  four-dimensional Abelian gravity}},  {\sl Phys. Lett.} {\bf B655} (2007)
  178--182, [\href{http://arxiv.org/abs/gr-qc/0702148}{{\sf
  arXiv:gr-qc/0702148}}],
  [\href{http://dx.doi.org/10.1016/j.physletb.2007.08.067}{{\sf
  doi:10.1016/j.physletb.2007.08.067}}].

\bibitem{Gao:2000ga}
S.~Gao and R.~M. Wald, {\it {Theorems on gravitational time delay and related
  issues}},  {\sl Class. Quant. Grav.} {\bf 17} (2000) 4999--5008,
  [\href{http://arxiv.org/abs/gr-qc/0007021}{{\sf arXiv:gr-qc/0007021}}],
  [\href{http://dx.doi.org/10.1088/0264-9381/17/24/305}{{\sf
  doi:10.1088/0264-9381/17/24/305}}].

\bibitem{Camanho:2014apa}
X.~O. Camanho, J.~D. Edelstein, J.~Maldacena, and A.~Zhiboedov, {\it {Causality
  Constraints on Corrections to the Graviton Three-Point Coupling}},  {\sl
  JHEP} {\bf 02} (2016) 020, [\href{http://arxiv.org/abs/1407.5597}{{\sf
  arXiv:1407.5597}}], [\href{http://dx.doi.org/10.1007/JHEP02(2016)020}{{\sf
  doi:10.1007/JHEP02(2016)020}}].

\bibitem{Hertzberg:2016djj}
M.~P. Hertzberg, {\it {Gravitation, Causality, and Quantum Consistency}},
  \href{http://arxiv.org/abs/1610.03065}{{\sf arXiv:1610.03065}}.

\bibitem{Hertzberg:2017abn}
M.~P. Hertzberg and M.~Sandora, {\it {General Relativity from Causality}},
  \href{http://arxiv.org/abs/1702.07720}{{\sf arXiv:1702.07720}}.

\bibitem{Zhang:2001xs}
S.-C. Zhang and J.-p. Hu, {\it {A Four-dimensional generalization of the
  quantum Hall effect}},  {\sl Science} {\bf 294} (2001) 823,
  [\href{http://arxiv.org/abs/cond-mat/0110572}{{\sf arXiv:cond-mat/0110572}}],
  [\href{http://dx.doi.org/10.1126/science.294.5543.823}{{\sf
  doi:10.1126/science.294.5543.823}}].

\bibitem{Gu:2009jh}
Z.-C. Gu and X.-G. Wen, {\it {Emergence of helicity +- 2 modes (gravitons) from
  qbit models}},  {\sl Nucl. Phys.} {\bf B863} (2012) 90--129,
  [\href{http://arxiv.org/abs/0907.1203}{{\sf arXiv:0907.1203}}],
  [\href{http://dx.doi.org/10.1016/j.nuclphysb.2012.05.010}{{\sf
  doi:10.1016/j.nuclphysb.2012.05.010}}].

\bibitem{Zaanen:2011hm}
J.~Zaanen and A.~J. Beekman, {\it {The Emergence of gauge invariance: The
  Stay-at-home gauge versus local-global duality}},  {\sl Annals Phys.} {\bf
  327} (2012) 1146--1161, [\href{http://arxiv.org/abs/1108.2791}{{\sf
  arXiv:1108.2791}}], [\href{http://dx.doi.org/10.1016/j.aop.2011.11.006}{{\sf
  doi:10.1016/j.aop.2011.11.006}}].

\bibitem{Price:2015qca}
H.~M. Price, O.~Zilberberg, T.~Ozawa, I.~Carusotto, and N.~Goldman, {\it
  {Four-Dimensional Quantum Hall Effect with Ultracold Atoms}},  {\sl Phys.
  Rev. Lett.} {\bf 115} (2015), no.~19 195303,
  [\href{http://arxiv.org/abs/1505.04387}{{\sf arXiv:1505.04387}}],
  [\href{http://dx.doi.org/10.1103/PhysRevLett.115.195303}{{\sf
  doi:10.1103/PhysRevLett.115.195303}}].

\bibitem{Ozawa:2016htj}
T.~Ozawa, H.~M. Price, N.~Goldman, O.~Zilberberg, and I.~Carusotto, {\it
  {Synthetic dimensions in integrated photonics: From optical isolation to
  four-dimensional quantum Hall physics}},  {\sl Phys. Rev.} {\bf A93} (2016),
  no.~4 043827, [\href{http://dx.doi.org/10.1103/PhysRevA.93.043827}{{\sf
  doi:10.1103/PhysRevA.93.043827}}].

\bibitem{Deffayet:2013tca}
C.~Deffayet, A.~E. Gümrükçüoğlu, S.~Mukohyama, and Y.~Wang, {\it {A no-go
  theorem for generalized vector Galileons on flat spacetime}},  {\sl JHEP}
  {\bf 04} (2014) 082, [\href{http://arxiv.org/abs/1312.6690}{{\sf
  arXiv:1312.6690}}], [\href{http://dx.doi.org/10.1007/JHEP04(2014)082}{{\sf
  doi:10.1007/JHEP04(2014)082}}].

\bibitem{Lovelock:1969uni}
D.~Lovelock, {\it {The uniqueness of the Einstein field equations in a
  four-dimensional space}},  {\sl Arch. Rational Mech. Anal.} {\bf 33} (1969)
  54--70, [\href{http://dx.doi.org/10.1007/BF00248156}{{\sf
  doi:10.1007/BF00248156}}].

\bibitem{Gao:2014jja}
X.~Gao, {\it {Derivative interactions for a spin-2 field at cubic order}},
  {\sl Phys. Rev.} {\bf D90} (2014), no.~6 064024,
  [\href{http://arxiv.org/abs/1403.6781}{{\sf arXiv:1403.6781}}],
  [\href{http://dx.doi.org/10.1103/PhysRevD.90.064024}{{\sf
  doi:10.1103/PhysRevD.90.064024}}].

\bibitem{Nutma:2013zea}
T.~Nutma, {\it {xTras : A field-theory inspired xAct package for mathematica}},
   {\sl Comput. Phys. Commun.} {\bf 185} (2014) 1719--1738,
  [\href{http://arxiv.org/abs/1308.3493}{{\sf arXiv:1308.3493}}],
  [\href{http://dx.doi.org/10.1016/j.cpc.2014.02.006}{{\sf
  doi:10.1016/j.cpc.2014.02.006}}].

\end{thebibliography}\endgroup

\end{document}